\documentstyle[12pt,psfig]{article}
\newcommand{\be}{\begin{equation}}
\newcommand{\ee}{\end{equation}}
\newcommand{\D}{\Delta}
\newcommand{\bea}{\begin{eqnarray}}
\newcommand{\eea}{\end{eqnarray}}

\begin{document}
\pagenumbering{arabic}
\title{Relativistic transport theory \\ for $N$, $\D$ and $N^{*}$(1440) system}
\author{
{\bf \underline{Guangjun Mao}, L. Neise, H. St\"ocker, W. Greiner}
\\[0.3cm]
{\small Institut f\"ur Theoretische Physik der J.W.Goethe Universit\"at}\\
{\small Postfach 111932, D-60054 Frankfurt a.M., Germany}
}
 \date{}
\maketitle   
\begin{abstract}
\begin{sloppypar}
A self-consistent relativistic  
Boltzmann-Uehling-Uhlenbeck equation  for
the $N^{*}$(1440) resonance is developed based on an effective Lagrangian
of baryons interacting through mesons. 
 The   equation 
is consistent with that of nucleon's and delta's which we 
derived before. Thus, we obtain a set of coupled equations for the $N$,
$\Delta$ and $N^{*}$(1440) distribution functions. 
 All the $N^{*}$(1440)-relevant in-medium two-body scattering
cross sections within the $N$, $\Delta$ and $N^{*}$(1440) system are derived
from the same effective Lagrangian in addition to the mean field and presented
analytically.  
Medium effects on the cross sections are discussed.
 \end{sloppypar}
\end{abstract}
\vspace{0.5cm}
\noindent {\bf I. Introduction}
\vspace{0.5cm}
\begin{sloppypar}
It was recognized twenty years ago that particles emitted in the collisions 
contain important information about the equation of state  
of hot and dense nuclear matter.
Since most of the particles such as pion, kaon, dilepton, anti-proton, 
anti-kaon are mainly produced through resonances, the inclusion of resonance
degrees of freedom in transport theories is essential for any realistic
description of relativistic heavy-ion collisions.
 Recent theoretical calculations \cite{Ehe93,Hof95} and experimental data 
\cite{Met93} indicated that a gradual transition to {\em resonance matter}
would occur in the collision zone at kinetic energy ranging from SIS 
($\sim$ 1AGeV) up to
AGS ($\sim$ 15AGeV). At an incident energy of 2 GeV/nucleon more than 30\% of 
the nucleons 
are excited to resonance states \cite{Ave94}. 
At intermediate- and high-energy range the most important 
baryonic resonances are $\Delta$(1232), $N^{*}$(1440) and $N^{*}$(1535). 
Theoretical models extended to describe relativistic heavy-ion collisions at 
this energy range should include these resonance degrees of freedom explicitly
and treat them self-consistently. The heart of the problem is to determine 
quantitatively all possible binary collisions relating to  resonances,
such as $N\Delta$, $\Delta\Delta$, $NN^{*}$(1440),
$NN^{*}$(1535) ... collisions.
Unfortunately, very little is known about resonance-relevant in-medium cross
sections in high-density nuclear matter since the experimental determination
of them is inaccessible yet. 

\end{sloppypar}
\begin{sloppypar}
In this contribution, we will derive the self-consistent RBUU equation
for the $N^{*}$(1440) distribution function within the framework we have done
for the nucleon's and $\Delta$'s. Special attentions will be paid to the 
$N^{*}$(1440)-relevant in-medium cross sections. Through construction the collision
term of $N^{*}$(1440)'s RBUU equation we will give the analytically expressions
for calculating all the $N^{*}$(1440)-relevant in-medium two-body scattering
cross sections within
the $N$, $\Delta$ and $N^{*}$(1440) system. The presented cross sections are 
consistent with the other ingredients of the transport model and can be used
directly in the study of relativistic heavy-ion collisions. 

\end{sloppypar}
 \vspace{0.5cm}
\noindent
{\bf II. RBUU-type transport equation for the $N^{*}$(1440) distribution
 function}
 \vspace{0.5cm}
\begin{sloppypar}
The effective Lagrangian which considers
the $N$, $\Delta$ and $N^{*}(1440)$ system interacting through $\sigma$,
$\omega$ and $\pi$ mesons can be written as
\begin{eqnarray}
 {\cal L}_{I}
 &=& {\rm g}_{NN}^{A}\bar{\psi}(x)\Gamma_{A}^{N} \psi(x) \Phi_{A}(x)
 + {\rm g}_{N^{*}N^{*}}^{A}\bar{\psi}^{*}(x)\Gamma_{A}^{N^{*}} \psi^{*}(x) \Phi_{A}(x)
 + {\rm g}_{\Delta \Delta}^{A} \bar{\psi}_{\Delta \nu}(x)\Gamma_{A}^{\Delta}\psi
 ^{\nu}_{\Delta}(x) \Phi_{A}(x) \nonumber \\
 && +\lbrack {\rm g}_{NN^{*}}^{A}\bar{\psi}^{*}(x)\Gamma_{A}^{N^{*}} \psi(x) \Phi_{A}(x)
 - {\rm g}_{\Delta N}^{\pi}\bar{\psi}_{\Delta \mu}(x)
 \partial^{\mu}\mbox{\boldmath $\pi$}(x) \cdot {\bf S}^{+}\psi(x) \nonumber \\
  && - {\rm g}_{\Delta N^{*}}^{\pi}\bar{\psi}_{\Delta \mu}(x)
 \partial^{\mu}\mbox{\boldmath $\pi$}(x) \cdot {\bf S}^{+}\psi^{*}(x) 
  + h.\, c. \rbrack
   \end{eqnarray}
where 
 $\psi_{\Delta \mu}$ is the Rarita-Schwinger spinor of the
$\Delta$-baryon.
 $\Gamma_{A}^{N}=\Gamma_{A}^{N^{*}}=
 \gamma_{A}\tau_{A}$, $\Gamma_{A}^{\Delta}=\gamma_{A}T_{A}$,
  A=$\sigma$, $\omega$, $\pi$;
${\rm g}_{NN}^{\pi}=f_{\pi}/m_{\pi}$,
 ${\rm g}_{\Delta N}^{\pi}=f^{*}/m_{\pi}$,
 the symbols and notation are defined in Table I.
  \begin{table}
     \vspace{-0.5cm}
    \begin{center}
    \tabcolsep 0.08in
  \begin{tabular}{|c|c|c|c|c|c|c|c|c|c|c|c|}
      \hline
 {\rm A} & m$_{A}$ & g$^{A}_{NN}$ & g$^{A}_{N^{*}N^{*}}$ & 
 g$^{A}_{\Delta \Delta}$ & g$^{A}_{NN^{*}}$ & $\gamma_{A} $ & $\tau_{A}$ & $ T_{A} $ &
  $\Phi_{A} (x) $ & D$_{A}^{\mu}$ & D$_{A}^{i}$ \\
 \hline
 $\sigma$ & m$_{\sigma}$ & g$^{\sigma}_{NN}$ & g$^{\sigma}_{N^{*}N^{*}}$ &
 g$_{\Delta \Delta}^{\sigma}$ & g$^{\sigma}_{NN^{*}}$ & 1 & 1 & 1 & $\sigma(x)$ &
 1 & 1   \\
 \hline
$\omega$ & m$_{\omega}$ & $-$ g$^{\omega}_{NN} $ & $-$ g$^{\omega}_{N^{*}N^{*}}$ & 
 $-$ g$_{\Delta \Delta}^{\omega}$ & $-$ g$^{\omega}_{NN^{*}}$ &  $\gamma_{\mu} $ &
 1 & 1 & $\omega^{\mu}(x)$ & $-$ g$^{\mu \nu}$ & 1 \\
 \hline
 $\pi$ & m$_{\pi}$ & g$_{NN}^{\pi}$& g$^{\pi}_{N^{*}N^{*}}$ & 
 g$_{\Delta \Delta}^{\pi}$ & g$^{\pi}_{NN^{*}} $ & 
 $\not\! k \gamma_{5} $
& \mbox{\boldmath $\tau$} &${\bf T}$
 & \mbox{\boldmath $\pi$}(x) & 1 & $\delta_{ij}$ \\
       \hline
      \end{tabular}
          \end{center}
\caption{Some symbols and notation used in this paper, k$_{\mu}$ is the 
 transformed four-momentum.}
 \end{table}

In the language of the closed time-path Green's function technique the 
$N^{*}$(1440) Green's function in the interaction picture                 
 can be defined in the same way as for nucleon's by
 \begin{equation}
iG_{N^{*}}(1,2)=<T[exp(-i \not \!\!\int\,dx H_{I}(x))\psi^{*}(1)\bar{\psi}^{*}(2)]> .
 \end{equation}
Here T is the time ordering operator defined on a time contour. 
The corresponding Dyson equation for $iG_{N^{*}}(1,2)$ can be written as
   \begin{equation}
 iG_{N^{*}}(1,2)=iG_{N^{*}}^{0}(1,2)+\not\!\! \int\,dx_{3} \not\!\! \int\,dx_{4}G_{N^{*}}^{0}(1,4)\Sigma_{N^{*}}(4,3)iG_{N^{*}}(3,2).
     \end{equation}
Here $G_{N^{*}}^{0}(1,2)$ is
the zeroth-order Green's function of $N^{*}$(1440), which 
is similar to that of the nucleon's
zeroth-order Green's function \cite{PRC96,PRC94}.
The only difference between the nucleon and the $N^{*}$(1440) is the 
mass and the coupling strengths! 
As in most/all presently used RBUU-type transport models, also here we do not 
take into account
the temperature degree of freedom. Furthermore,
in our theoretical framework the negative-energy states are neglected.
$\Sigma_{N^{*}}(4,3)$  is the $N^{*}$ self-energy. The lowest-order 
self-energies contributing to the collision term come from the Born diagrams.
Through considering the $N^{*}$ self-energy up to the Born approximation and 
adopting the semi-classical approximation 
 and quasi-particle approximation 
  the self-consistent
RBUU equation for the $N^{*}$(1440) can be derived in the same way as that
of the nucleons.
 The RBUU equation for the $N^{*}$(1440)
distribution function reads 
 \begin{equation}
\lbrace p_{\mu} [
 \partial^{\mu}_{x}-\partial^{\mu}_{x}\Sigma_{N^{*}}
^{\nu}(x) \partial_{\nu}^{p}+\partial_{x}^{\nu}\Sigma_{N^{*}}^{\mu}(x)\partial
_{\nu}^{p} ] +m^{*}_{N^{*}}\partial^{\nu}_{x}\Sigma_{N^{*}}^{S}(x)\partial_{\nu}^{p}
\rbrace \frac{f_{N^{*}}({\bf x}, {\bf p}, \tau)}{E^{*}_{N^{*}}(p)}
  = C^{N^{*}}(x,p),
 \end{equation}
where $f_{N^{*}}({\bf x}, {\bf p}, \tau)$ is the single-particle distribution
function of the $N^{*}(1440)$.
The left-hand side of Eq. (4) is the transport part and the right-hand side
is the collision term. Here we have dropped the contribution of the Fock term,
since it usually has only small effects on the mean field. 
The above equation is derived within the framework as we have done for 
the nucleon's \cite{PRC94,PRC97} 
 \begin{equation}
\lbrace p_{\mu} [
 \partial^{\mu}_{x}-\partial^{\mu}_{x}\Sigma
^{\nu}(x) \partial_{\nu}^{p}+\partial_{x}^{\nu}\Sigma^{\mu}(x)\partial
_{\nu}^{p} ] +m^{*}\partial^{\nu}_{x}\Sigma^{S}(x)\partial_{\nu}^{p}
\rbrace \frac{f({\bf x}, {\bf p}, \tau)}{E^{*}(p)}
  = C(x,p).
 \end{equation}
and delta's \cite{PRC96}
 \begin{equation}
\lbrace p_{\mu} [
 \partial^{\mu}_{x}-\partial^{\mu}_{x}\Sigma_{\Delta}
^{\nu}(x) \partial_{\nu}^{p}+\partial_{x}^{\nu}\Sigma_{\Delta}^{\mu}(x)\partial
_{\nu}^{p} ] +m^{*}_{\Delta}\partial^{\nu}_{x}\Sigma_{\Delta}^{S}(x)\partial_{\nu}^{p}
\rbrace \frac{f_{\Delta}({\bf x}, {\bf p}, \tau)}{E^{*}_{\Delta}(p)}
  = C^{\Delta}(x,p).
 \end{equation}
RBUU equations. Therefore, Eqs. (4), (5) and (6) stand in a consistent 
form and they are coupled together through the mean field and collision
term(i.e., in-medium scattering cross sections of different channels).  
The $\Sigma^{S}_{N^{*}}(x)$ and $\Sigma^{\mu}_{N^{*}}(x)$ are the Hartree
terms of the scalar and vector $N^{*}$(1440) self-energies. 
After taking into account the self-interaction of the $\sigma$, $\omega$ fields
 the field equations of the $\sigma$ and $\omega$ mesons can be written as
 \begin{eqnarray}
 m_{\sigma}^{2}\sigma(x)+b({\rm g}^{\sigma}_{N N})^{3}\sigma^{2}(x)
+c({\rm g}^{\sigma}_{N N})^{4}\sigma^{3}(x)={\rm g}^{\sigma}_{N N}\rho_{S}(N)
+{\rm g}^{\sigma}_{N^{*} N^{*}}\rho_{S}(N^{*}) +{\rm g}^{\sigma}_{\Delta \Delta}\rho_{S}(\Delta), \\
 m_{\omega}^{2}\omega^{\mu}(x)+\frac{({\rm g}_{N N}^{\omega})^{2}m_{\omega}^{2}}
{Z^{2}}(\omega^{\mu}(x))^{3}
={\rm g}^{\omega}_{N N}\rho_{V}^{\mu}(N)+{\rm g}^{\omega}_{N^{*} N^{*}}\rho_{V}^{\mu}
 (N^{*}) + {\rm g}^{\omega}_{\Delta \Delta}\rho_{V}^{\mu}(\Delta).
 \end{eqnarray}
The effective four momentum and effective mass of the $N^{*}$(1440) are defined as 
  \begin{eqnarray}
    &&m^{*}_{N^{*}}(x)=M_{N^{*}} - {\rm g}^{\sigma}_{N^{*} N^{*}}\sigma(x) \\
    &&p^{\mu}(x)=P^{\mu} - {\rm g}^{\omega}_{N^{*} N^{*}} \omega^{\mu}(x).
  \end{eqnarray}
Here $\rho_{S}$(i) and $\rho_{V}^{\mu}$(i) are the scalar and vector densities
of the nucleon, $N^{*}$(1440) and delta:
 \begin{eqnarray}
&& \rho_{S}(i)=\frac{\gamma(i)}{(2 \pi)^{3}}\int d{\bf q} \frac{m^{*}_{i}}
 {\sqrt{{\bf q}^{2}+m^{*2}_{i}}} f_{i}({\bf x},{\bf q},\tau), \\
&& \rho_{V}^{\mu}(i)=\frac{\gamma(i)}{(2 \pi)^{3}}\int d{\bf q} \frac{q^{\mu}}
 {\sqrt{{\bf q}^{2}+m^{*2}_{i}}} f_{i}({\bf x},{\bf q},\tau).
 \end{eqnarray}
The abbreviations i=N, $N^{*}$, $\Delta$, and $\gamma$(i)= 4, 4, 16, correspond 
 to nucleon, $N^{*}$(1440) and delta, respectively.   
 The effective four-momenta and effective masses of 
nucleon and delta can be defined through substituting the appropriate nucleon
and delta labels in Eqs. (9) and (10), respectively.  

  \end{sloppypar}
  \begin{sloppypar}
The collision term can be expressed according to the transition probability,
which reads as
   \begin{eqnarray}
  C^{N^{*}}(x,p)&=&\frac{1}{2}\int \frac{d^{3}p_{2}}{(2\pi)^{3}}
 \int\frac{d^{3}p_{3}}{(2\pi)^{3}} \int\frac{d^{3}p_{4}}{(2\pi)^{3}}
 (2\pi)^{4}\delta^{(4)}(p+p_{2}-p_{3}-p_{4})\nonumber \\
 && \times W^{N^{*}}(p,\,p_{2},\,p_{3},\,p_{4})
 (F_{2}-F_{1}),
    \end{eqnarray}
where $F_{2}$, $F_{1}$ are the Uehling-Uhlenbeck factors of the gain 
($F_{2}$) and loss ($F_{1}$) terms, respectively:
    \begin{eqnarray}
&& F_{2}=[1-f_{N^{*}}({\bf x},{\bf p},\tau)][1-f_{B_{2}}({\bf x},{\bf p}_{2},\tau)]
   f_{B_{3}}({\bf x},{\bf p}_{3},\tau)f_{B_{4}}({\bf x},{\bf p}_{4},\tau), \\
&&  F_{1}=f_{N^{*}}({\bf x},{\bf p},\tau)f_{B_{2}}({\bf x},{\bf p}_{2},\tau)
   [1-f_{B_{3}}({\bf x},{\bf p}_{3},\tau)]
   [1-f_{B_{4}}({\bf x},{\bf p}_{4},\tau)] , 
    \end{eqnarray}
$B_{2}$, $B_{3}$, $B_{4}$ can be $N$, $\Delta$ and $N^{*}$(1440).
$W^{N^{*}}(p,\,p_{2},\,p_{3},\,p_{4})$  is the transition
 probability of different channels, which has the form
  \begin{equation}
 W^{N^{*}}(p,\,p_{2},\,p_{3}\,,p_{4})=
  \frac{1}{16E^{*}_{N^{*}}(p)E^{*}_{B_{2}}(p_{2})E^{*}_{B_{3}}(p_{3})
  E^{*}_{B_{4}}(p_{4})}\sum_{AB} (T_{D}\Phi_{D} - T_{E}\Phi_{E})
   +p_{3} \longleftrightarrow p_{4}.
  \end{equation}
Here $T_{D}$, $T_{E}$ are the isospin matrices and $\Phi_{D}$, $\Phi_{E}$
are the spin matrices, respectively.
 D denotes the contribution of the direct diagrams and E is that of the exchange
diagrams. $A$, $B=\sigma$, $\omega$, $\pi$ represent the contributions of 
different mesons. The exchange of $p_{3}$ and $p_{4}$ is only for the 
case of identical particles in the final state. 
 The two-body scattering reactions relevant to the $N^{*}$(1440) 
in the $N$, $\Delta$ and $N^{*}$(1440) system are follows: \\
 \indent (1) Elastic reactions: \\
\mbox{} \hspace{1.0cm} $NN^{*} \longrightarrow NN^{*}$, \hspace{1.0cm}
  $\Delta N^{*} \longrightarrow \Delta N^{*}$, \hspace{1.0cm} 
  $N^{*}N^{*} \longrightarrow N^{*} N^{*}$ .   \\
  \indent (2) Inelastic reactions:  \\
 \mbox{} \hspace{1.0cm} $N N \longleftrightarrow NN^{*}$, \hspace{1.0cm} 
  $N \Delta \longleftrightarrow N N^{*}$, \hspace{1.0cm} 
  $\Delta \Delta \longleftrightarrow N N^{*}$, \\
 \mbox{} \hspace{1.0cm} $N N^{*} \longleftrightarrow \Delta N^{*}$, \hspace{0.9cm} 
  $N N^{*} \longleftrightarrow N^{*} N^{*}$, \hspace{0.6cm} 
  $N N \longleftrightarrow \Delta N^{*}$, \\
 \mbox{} \hspace{1.0cm} $N \Delta \longleftrightarrow \Delta N^{*}$,
\hspace{1.0cm} $\Delta \Delta \longleftrightarrow \Delta N^{*}$, \hspace{1.0cm} 
  $N^{*} N^{*} \longleftrightarrow \Delta N^{*}$ , \\
 \mbox{} \hspace{1.0cm} $N N \longleftrightarrow N^{*}N^{*}$, \hspace{0.8cm} 
  $N \Delta \longleftrightarrow N^{*}N^{*}$, \hspace{0.8cm}
  $\Delta \Delta \longleftrightarrow N^{*}N^{*}$. \\
\noindent For the inelastic case    we only 
calculate the $N^{*}$(1440)-incident cross sections, its vice versa cross
sections can be obtained by means of the detailed balance \cite{Dan91}.
We have derived the analytical expressions for the above differential cross
sections through calculating the Born term of the $N^{*}$(1440) self-energies.
By means of the relation between the transition probability $W^{N^{*}}(p, p_{2},
p_{3}, p_{4})$ and the differential cross section \cite{Gro80}, Eq. (13)
can be rewritten as
  \begin{equation}
 C^{N^{*}}(x,p)=\frac{1}{2}\int \, \frac{d^{3}p_{2}}{(2\pi)^{3}}
 \upsilon \sigma_{N^{*}}(s,t)(F_{2}-F_{1})d\,\Omega.
  \end{equation}
Here $\upsilon$ is the M$\not\!\! o$ller velocity, $ \sigma_{N^{*}}(s,t)$ is the
differential cross section of different $N^{*}$-incident channels. 
The concrete expression of $\sigma_{N^{*}}(s,t)$ is  given in Ref. \cite{PRC}. 

  \end{sloppypar} 
 \vspace{0.5cm}
\noindent
{\bf III. The centroid $N^{*}$(1440) mass, coupling strengths and form factors}
 \vspace{0.5cm}
  \begin{sloppypar}
In order to take into account the broad decay width of the $N^{*}$(1440)
resonance, we introduce a 
centroid $N^{*}$(1440) mass $\langle M_{N^{*}} \rangle$ in the same way as 
we did for the delta \cite{PRC94,PRC97}.
$\langle M_{N^{*}} \rangle $ is defined as
  \begin{equation}
  \langle M_{N^{*}} \rangle = \frac{\int_{M_{N}+m_{\pi}}^{\sqrt{S}-M_{N}}
 f(M_{N^{*}})M_{N^{*}} d \, M_{N^{*}} }
 {\int_{M_{N}+m_{\pi}}^{\sqrt{S}-M_{N}} f(M_{N^{*}}) d\, M_{N^{*}} },
  \end{equation}
$f(M_{N^{*}})$ is the Breit-Wigner function 
  \begin{equation}
 f(M_{N^{*}}) = \frac{1}{2\pi} \frac{\Gamma(q)}{(M_{N^{*}}-M_{0})^{2}
 +\frac{1}{4}\Gamma^{2}(q) },
  \end{equation}
here $M_{0}=1440$ MeV and $\Gamma(q)$ is the momentum-dependent decay width 
 \cite{PhD}
 \begin{equation}
 \Gamma(q)=\Gamma_{0} \frac{M_{0}}{M_{N^{*}}}(q/q_{0})^{3} 
 \frac{1.2}{1+0.2(\frac{q}{q_{0}})^{2}}, 
 \end{equation}
$q_{0}$ is related to the case of $M_{N^{*}}=M_{0}$ and $\Gamma_{0}=200$ MeV.
 The effect of the decay
width of $N^{*}$(1440) is taken into account through replacing $M_{N^{*}}$
in Eq. (9) with $ \langle M_{N^{*}} \rangle $. The in-medium 
$N^{*} + N \rightarrow N+ N$ and $N^{*} + N \rightarrow N^{*} + N$ cross
sections can then be calculated by means of the equations
 \begin{eqnarray}
 &&\sigma^{*}_{N^{*}N \rightarrow NN} = \frac{1}{16N} \int \sigma_{N^{*}N
 \rightarrow NN} (s,t) d\, \Omega, \\
 &&\sigma^{*}_{N^{*}N \rightarrow N^{*}N} = \frac{1}{8} \int \sigma_{N^{*}N
 \rightarrow N^{*}N }(s,t) d\, \Omega,
 \end{eqnarray}
here $N$ is the normalization factor stemming from the decay width of the
$N^{*}$(1440) \cite{Dan91,Wol92,PRC97}.
The in-medium $N^{*}$(1440) production cross section can be obtained from 
Eq. (21) through detailed balance \cite{Dan91}
 \begin{equation}
\sigma^{*}_{NN \rightarrow NN^{*}} = \frac{1}{8} \int \frac{p^{2}_{NN^{*}}}
 {p^{2}_{NN}} \sigma_{N^{*}N \rightarrow NN} (s,t) d\, \Omega , 
 \end{equation}
where $p_{NN}$, $p_{NN^{*}}$ denote the c. m. three momentum of the {\em NN}
and $NN^{*}$ states, respectively.  

  \end{sloppypar}
  \begin{sloppypar}
For the coupling strength of 
${\rm g}_{NN}^{\pi}$, we take the most commonly used value $f^{2}_{\pi}
 /4\pi=0.08$. The coupling strengths of ${\rm g}_{NN}^{\sigma}$ and
${\rm g}_{NN}^{\omega}$ are determined by fitting the known ground-state
properties for infinite nuclear matter. 
For the coupling strengths of nucleon-$N^{*}$(1440) coupling we follow the 
arguments of Ref. \cite{Hub94}. The following relation is expected to be
valid
 \begin{equation}
\frac{{\rm g}_{NN^{*}}^{\pi}}{{\rm g}_{NN}^{\pi}}=
\frac{{\rm g}_{NN^{*}}^{\sigma}}{{\rm g}_{NN}^{\sigma}}=
\frac{{\rm g}_{NN^{*}}^{\omega}}{{\rm g}_{NN}^{\omega}} .
 \end{equation}
${\rm g}_{NN^{*}}^{\pi}$ is determined from the width of pion decay of the
$N^{*}$(1440)-resonance
$ {\rm g}_{NN^{*}}^{\pi}/{\rm g}_{NN}^{\pi}=0.351$.
 For the $N^{*}N^{*}$ coupling strengths, unfortunately, there is no  
information directly  available from experiments. A similar situation 
exists for the $\Delta\Delta$ coupling strengths. Based on the quark model
and mass splitting arguments several different choices for the delta coupling
strengths have been discussed in Refs. \cite{Mos74,Wal87}, which will cause
strong influence on the nuclear equation of state in relativistic mean field
calculations \cite{Wal87}.  
We assume that the arguments of 
Refs. \cite{Mos74,Wal87} apply to the
$N^{*}N^{*}$ coupling strengths and mainly consider the following three cases:
 \begin{equation}
\alpha(N^{*})=\frac{{\rm g}_{N^{*}N^{*}}^{\omega}}
{{\rm g}_{NN}^{\omega}}=1, \hspace{2cm} 
\beta(N^{*})=\frac{{\rm g}_{N^{*}N^{*}}^{\sigma}}
{{\rm g}_{NN}^{\sigma}}=1, 
 \end{equation}
 \begin{equation}
 \alpha(N^{*})=\beta(N^{*})=\frac{M_{N^{*}}}{M_{N}} \approx 1.5.
 \end{equation}
and
 \begin{equation}
 \alpha(N^{*})=1, \hspace{2cm} \beta(N^{*}) \approx 1.5
 \end{equation}
The influence of different choices of  $\alpha(N^{*})$ and
$\beta(N^{*})$ on the predicted 
in-medium cross sections  
 will be checked. For simplicity, an universal coupling strength
of ${\rm g}_{\Delta\Delta}^{\pi}={\rm g}_{N^{*}N^{*}}^{\pi}={\rm g}_{NN}^{\pi}$
is always assumed.

 \end{sloppypar}
 \begin{sloppypar}
To take account of the effects stemming from the finite size of hadrons and a
 part of the short range correlation, a phenomenological form factor is
 introduced at each vertex.
Here we distinguish the form factor $\Lambda^{*}_{A}$ 
for the nucleon-$N^{*}$(1440)-meson vertex to the $\Lambda_{A}$ for the 
nucleon-nucleon-meson vertex. S.~Huber and J.~Aichelin claimed that 
$\Lambda^{*}_{A}$ is about 40\% of $\Lambda_{A}$ \cite{Hub94}. We adopt this
argument in the numerical calculations.
 The form factor of the $N^{*}$(1440)-$N^{*}$(1440)-meson vertex is taken
  the same as
that of corresponding nucleon-nucleon-meson vertex. The cut-off masses
 $\Lambda_{\sigma}$=1200 MeV,
$\Lambda_{\omega}$=808 MeV and  $\Lambda_{\pi}$=500 MeV fixed in Refs. 
\cite{PRC94,PRC97}
will be used,  which are obtained by fitting  the experimental data of nucleon
mean free path
and the free {\em NN} scattering cross section. According to the above
argument, $\Lambda^{*}_{\sigma}$=480 Mev, $\Lambda^{*}_{\omega}$=323 MeV.
But we still take $\Lambda^{*}_{\pi}=\Lambda_{\pi}$=500 MeV, since this value
is already comparable to the $\Lambda^{*}_{\pi}$=400 MeV used in 
Ref. \cite{Hub94}.

 \end{sloppypar}
 \vspace{0.5cm}
\noindent {\bf IV. Numerical results and discussions}
\begin{figure}[htbp]
 \vspace{-0.5cm}
\hskip 2.2cm \psfig{file=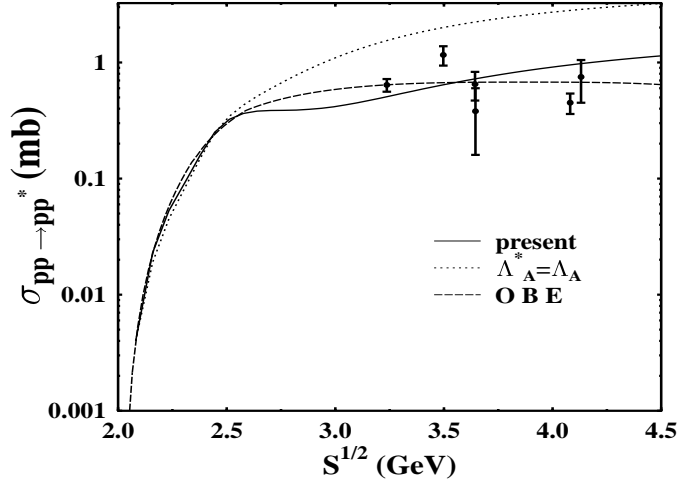,width=9cm,height=9cm,angle=-90}
 \vskip -0.5cm
\caption{Free scattering cross section for reaction $pp \rightarrow
 pp^{*}(1440)$. 
 Solid line represents the results of this work, and dashed
line denotes the results of Ref. \protect{\cite{Hub94}}.
 The experimental data are taken from Ref. \protect{\cite{CERN}}.
The unitary form factor for $NN$ and $NN^{*}$ vertex, i.e., $\Lambda^{*}_{A}
 =\Lambda_{A}$ is also tested in the calculations, which is depicted by the
 dotted line.}
\end{figure}
 \begin{sloppypar}
In Fig.~1 we compare our theoretical
predications 
of free $pp \rightarrow pp^{*}(1440)$ cross section to the available 
experimental data \cite{CERN}. The results of the one-boson-exchange model
computed by Huber and Aichelin \cite{Hub94} are also presented in this figure
as dashed line. Our results are  consistent with that of Ref. \cite{Hub94}.
Both of them are in good agreement with the experimental data. Here we should
point out that our calculations are almost parameter free. We do not fit any
parameters to the predicted cross section. Only the argument of $\Lambda^{*}
_{\sigma} / \Lambda_{\sigma} = \Lambda^{*}_{\omega} / \Lambda_{\omega}$=40\%
is taken from Ref. \cite{Hub94}. If $\Lambda^{*}_{\sigma}=\Lambda_{\sigma}$
and $\Lambda^{*}_{\omega}=\Lambda_{\omega}$ are adopted, the cross section will
be three times larger than the empirical value at higher energy as indicated
by the dotted line in the figure.

 \end{sloppypar}
\begin{minipage}{8.0cm}
  \hspace{-0.5cm}
 \centerline{\psfig{figure=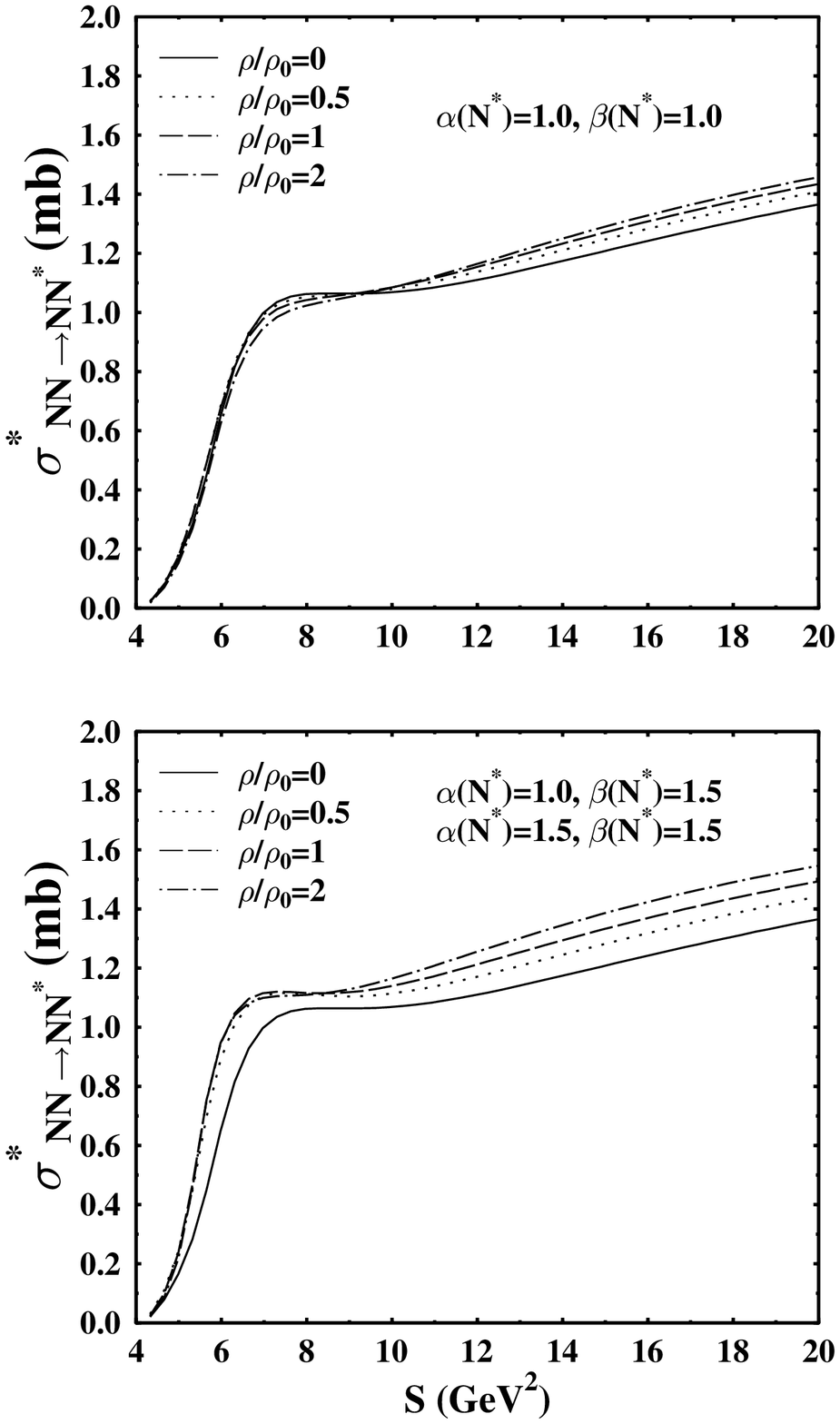,width=8.0cm,height=10.0cm,angle=0}}
 \end{minipage}\hfill
\begin{minipage}{8.0cm}
 Fig.~2: The in-medium $NN \rightarrow NN^{*}$ cross section at
different densities and energies. The calculations are performed with 
different sets of $\alpha(N^{*})$ and $\beta(N^{*})$.
 \end{minipage}
 \begin{sloppypar}
Fig.~2 displays the in-medium $N^{*}$(1440) production cross sections at 
different densities and energies. 
 It is shown from the figure that
the $\sigma^{*}_{NN \rightarrow NN^{*}}$ increases with the increase of
density. When  the universal coupling strengths are used, only a mild 
dependence on the density is exhibited. The density dependence, however, will
become evident if one uses a larger scalar-$N^{*}$(1440) coupling strength.
The choice of the $\alpha(N^{*})$ has no influence on the predicted cross 
sections. The reason is as follows: firstly, 
${\rm g}_{N^{*}N^{*}}^{\omega}$ does not enter the expressions of the $\sigma
_{NN \rightarrow NN^{*}}^{*}$ explicitly; secondly, we always calculate
the in-medium total energy of two particle system (small s) from the incident
two particles, i.e., two nucleons in the case of the $\sigma^{*}_{NN \rightarrow
 NN^{*}}$. The situation will change if
one considers the $\sigma_{N^{*}N \rightarrow NN}^{*}$, where the influence
of $\alpha(N^{*})$ will enter in the calculations of in-medium total energy
(small s) from free total energy (capital S), and then affects the in-medium
cross section.

 \end{sloppypar}
\begin{minipage}{8.0cm}
 \centerline{\psfig{figure=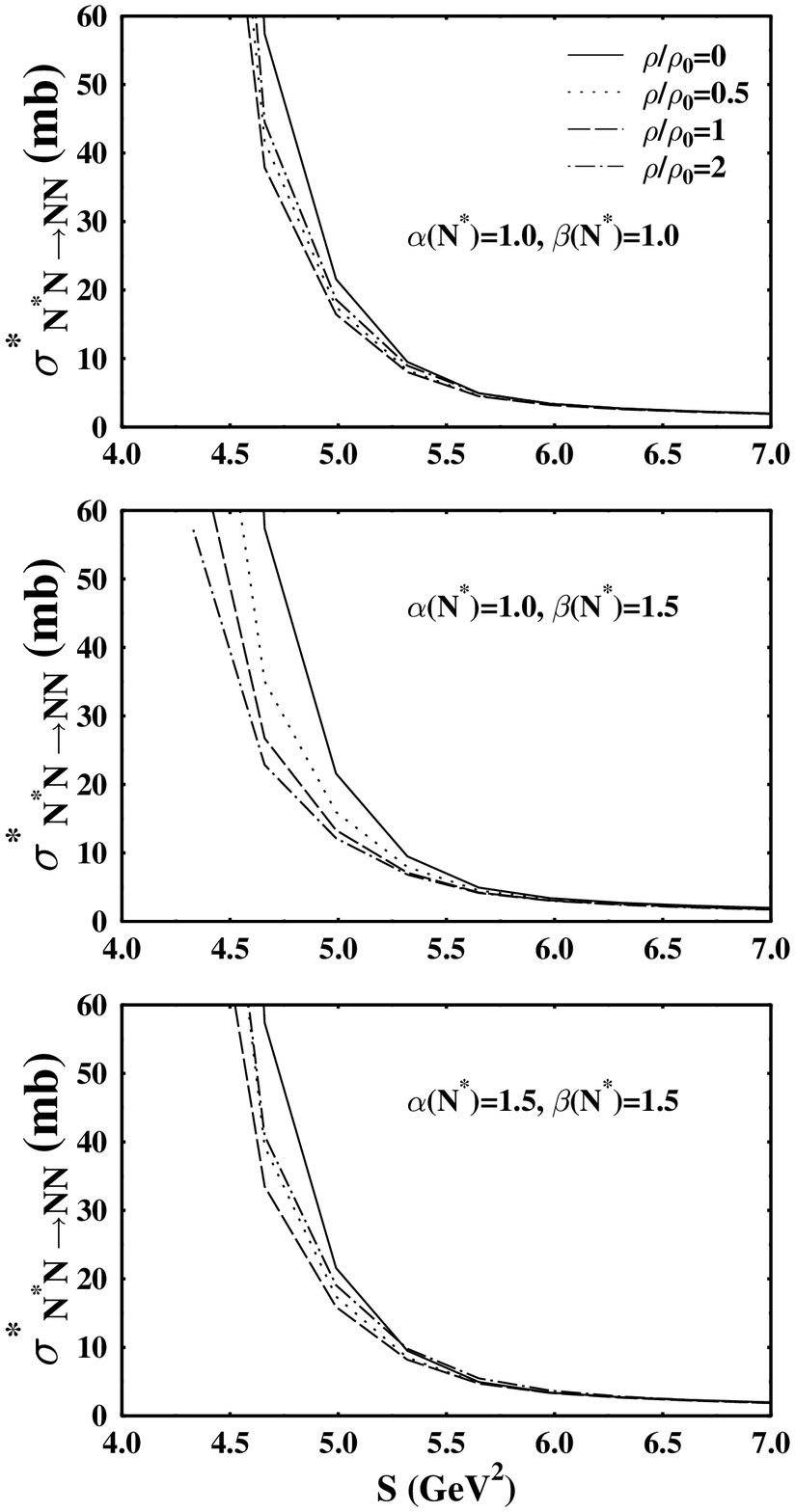,width=8.0cm,height=14.0cm,angle=0}}
 \vspace{-0.3cm}
 Fig.~3: The same as Fig.~2, but for an \\[-0.0cm] in-medium $N^{*} N
 \rightarrow NN$ cross section. 
 \end{minipage}\hfill
\begin{minipage}{8.0cm}
 \centerline{\psfig{figure=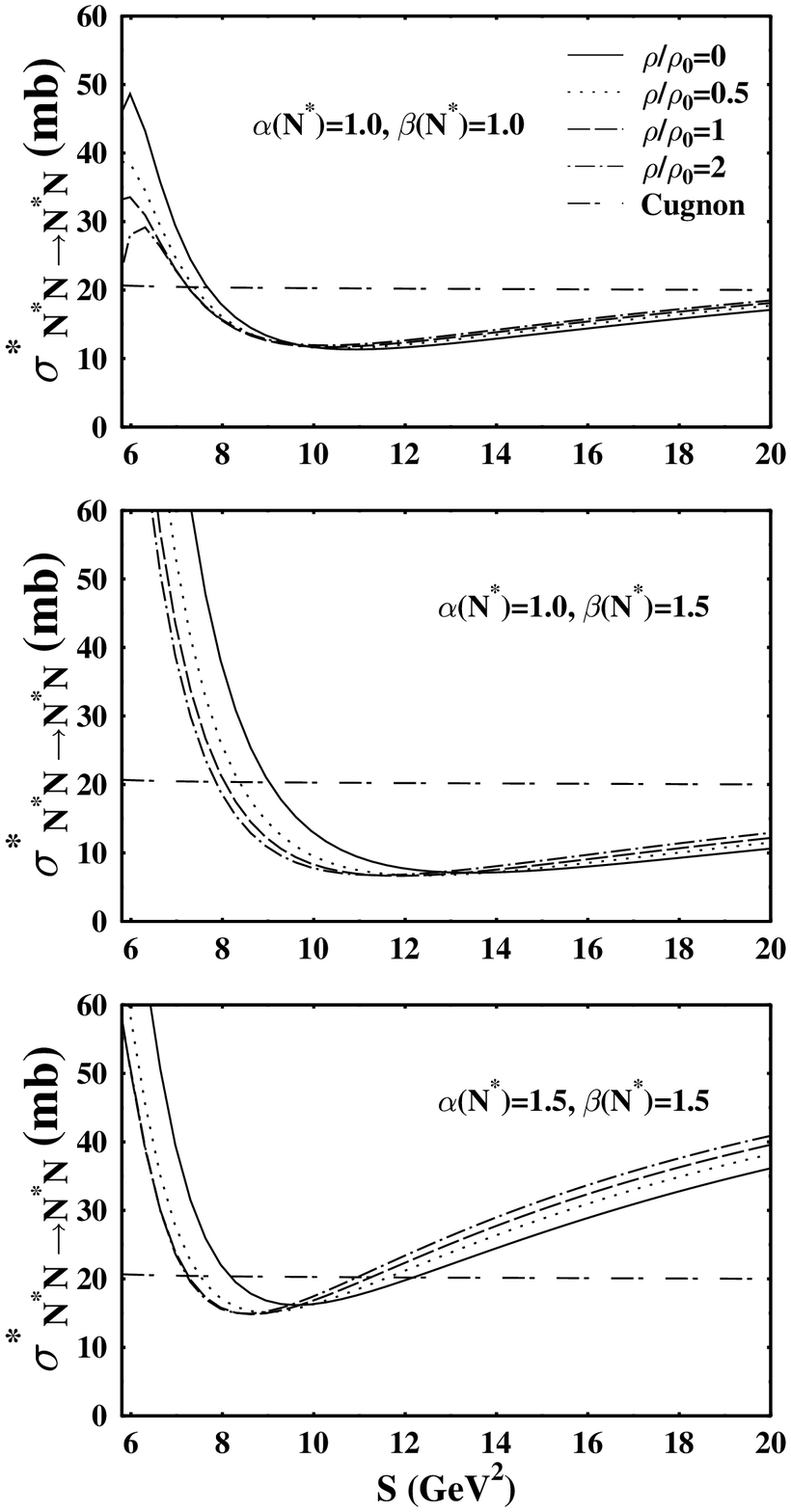,width=8.0cm,height=14.0cm,angle=0}}
 \vspace{-0.3cm}
 Fig.~4: The same as Fig.~2, but for an \\[-0.0cm] 
 in-medium $N^{*} N
 \rightarrow N^{*}N$ cross section.
 \end{minipage}
 \vspace{0.5cm}
 \begin{sloppypar}
Fig.~3 depicts the in-medium $N^{*}$(1440) absorption cross section. 
The cross sections drop 
very rapidly when the energy exceeds the threshold. That means that the
absorption process are most important at energy close to the threshold as in
the case of $\Delta$ absorption. When $\alpha(N^{*})=1$, $\beta(N^{*})=1.5$
is used as the $N^{*}$(1440) coupling strengths, the $\sigma_{N^{*}N \rightarrow
 NN}^{*}$ exhibits a clear
 density dependence. It decreases with the increase
 of density. In other two cases, i.e., $\alpha(N^{*})=\beta(N^{*})=1$ and
 $\alpha(N^{*})=\beta(N^{*})=1.5$, the dependence of the $\sigma_{N^{*}N
\rightarrow NN}^{*}$ on the density becomes weaker and less explicit.

 \end{sloppypar}
%\begin{figure}[htbp]
%\hskip 1.2cm \psfig{file=fig13.ps,width=7cm,height=10cm,angle=0}
%\vskip 0.4cm
%\end{figure}
 \begin{sloppypar}
 In Fig.~4 we show the in-medium $N^{*}N \rightarrow N^{*}N$ cross section at
different densities and energies. As can be found from the figure, the cross
sections now become very sensitive to the $\alpha(N^{*})$ and $\beta(N^{*})$
used because ${\rm g}_{N^{*}N^{*}}^{\sigma}$ and ${\rm g}_{N^{*}N^{*}}^{\omega}$
 enter the expressions of the $\sigma_{N^{*}N \rightarrow N^{*}N}^{*}$
explicitly. Generally speaking, the density dependence of the
cross section is not very evident when $\alpha(N^{*})=\beta(N^{*})=1$ and
$\alpha(N^{*})=\beta(N^{*})=1.5$ are used, mainly due to the strong cancelation
 effects from the $\sigma + \omega$ mixed term. A strong density
dependence appears when the set of $\alpha(N^{*})=1$, $\beta(N^{*})=1.5$ is 
used as the $N^{*}$(1440) coupling strengths. The in-medium cross section
decreases with the increase of density at lower energy and increases at higher
energy. As the energy changes, the cross section firstly decreases and then 
increases, especially in the case of $\alpha(N^{*})$=1.5. It is mainly caused 
by the contribution of the $\omega$ term. The
$\omega$ term approaches a saturation with the increase of energy while all
other terms (especially, the important cancelation term of the $\sigma + \omega
 $ mixed term) decrease. 
The Cugnon's parameterization \cite{Cug81} for free {\em NN}
elastic cross section, which is commonly used in the transport models for the 
$N^{*}N$ elastic cross section, is also plotted in Fig.~4. One can find an 
evident difference between  the
in-medium $N^{*} + N \rightarrow N^{*} + N$ cross section and the Cugnon's
parameterization. It is therefore important to take the in-medium cross sections
into account in the study of heavy-ion collisions.

 \end{sloppypar}
 \vspace{0.5cm}
\noindent {\bf V. Summary}
 \vspace{0.5cm}
  \begin{sloppypar}
Starting from the effective Lagrangian describing baryons interacting through 
 mesons, using the closed time-path Green's function technique and adopting
 the semi-classical, quasi-particle and Born approximations
we have developed a RBUU-type transport equation for the $N^{*}$(1440)
distribution function. The equation is derived within the same framework
which was successfully applied to the
nucleon \cite{PRC94} and delta \cite{PRC96} and thus we obtained
a set of self-consistent equations for the $N$, $\Delta$ and $N^{*}$(1440)
system. Three equations
are coupled through the self-energy terms and collision terms and should be
solved simultaneously in a numerical simulation of heavy-ion collisions. 
Both the mean field and collision term of the $N^{*}$(1440)'s RBUU equation are 
derived from the same effective Lagrangian and given explicitly, so the medium
effects on the two-body scattering cross sections are addressed automatically  
and can be studied 
self-consistently. Therefore, this approach
  provides a promising way to reach a covariant
description of the $N^{*}$(1440) in relativistic heavy-ion collisions. 

Based on this approach, we have studied 
the in-medium two-body scattering cross sections. Since there is no 
information about the $N^{*}N^{*}$ coupling strengths available, several
different choices for $\alpha(N^{*})={\rm g}_{N^{*}N^{*}}^{\omega} / {\rm g}
_{NN}^{\omega}$ and $\beta(N^{*}) = {\rm g}_{N^{*}N^{*}}^{\sigma} / {\rm g}
_{NN}^{\sigma}$ are investigated. The results turn out to be sensitive to the
$\alpha(N^{*})$ and $\beta(N^{*})$ used. 
Generally speaking,
only a mild density-dependence of in-medium cross sections is found in the
cases of $\alpha(N^{*})=\beta(N^{*})=1$ and $\alpha(N^{*})=\beta(N^{*})=1.5$.
The situation, however, is changed when the set of $\alpha(N^{*})=1$, $\beta
(N^{*})=1.5$ is adopted. An evident density-dependence  appears. 
Qualitatively, the $\sigma^{*}_{NN \rightarrow NN^{*}}$ are found to increase
with the increase of density while the $\sigma^{*}_{N^{*}N \rightarrow NN}$
near the threshold energy decreases.
For the $\sigma^{*}_{N^{*}N \rightarrow N^{*}N}$, the situation is a little 
complicated. It decreases with the increase of density at lower energy and 
increases at higher energy. 
Because we have not included the screening and anti-screening
effects of the medium on the interaction in the present calculations,
the above arguments should  be viewed with caution.  Further
 investigations are needed.

 \end{sloppypar}

\end{document}